\documentclass[notitlepage,a4paper,aps,prd,twocolumn,superscriptaddress,nofootinbib,groupedaddress]{revtex4}
\usepackage{graphicx}
\usepackage[colorlinks=true, pdfstartview=FitV, linkcolor=blue, citecolor=red, urlcolor=magenta]{hyperref}

\usepackage{dcolumn}
\usepackage{graphics}
\usepackage{amssymb}
\usepackage{amsmath}
\usepackage{url}
\usepackage{color}
\usepackage[utf8]{inputenc}

\usepackage{ulem}
\usepackage{tikz}
\usetikzlibrary{arrows,decorations.markings}

\newcommand{\rmd}{\textrm{d}}

\newcommand{\kit}{\textit{k}}
\begin{document}
\newcommand{\cbpf}{
\affiliation{Department of Cosmology, Astrophysics and Fundamental Interacions-COSMO, Centro Brasileiro de Pesquisas F\'{\i}sicas-CBPF, rua Dr. Xavier Sigaud 150, 22290-180, Rio de Janeiro, Brazil.}}
\newcommand{\cecs}{
\affiliation{Centro de Estudios Cient\'{\i}ficos (CECs), Av. Arturo Prat 514, Valdivia, Chile.}}

\newcommand{\uss}{
\affiliation{Universidad San Sebasti\'an, General Lagos 1163, Valdivia, Chile.}}

\newcommand{\ufes}{\affiliation{PPGCosmo, CCE - Universidade Federal do Esp\'{\i}rito Santo, Vit\'{o}ria-ES, 29075-910, Brazil.}}

\title{Inflation and late-time accelerated expansion driven by k-essence degenerate dynamics}

\author{Alexsandre L. Ferreira Junior}\email{alexsandre.ferreira@edu.ufes.br}
\ufes
\cecs
\author{Nelson Pinto-Neto}\email{nelsonpn@cbpf.br}
\cbpf
\author{Jorge Zanelli}\email{z@cecs.cl}
\cecs \uss

\date{\today}

\begin{abstract}
We consider a \kit--essence model in which a single scalar field can be responsible for both primordial inflation and the present observed acceleration of the cosmological background geometry, while also admitting a non-singular de Sitter beginning of the Universe (it arises from de Sitter and ends in de Sitter). The early one is driven by a slow roll potential, and the late one through a dynamical dimensional reduction process which freezes the scalar field in a degenerate surface, turning it into a cosmological constant. This is done by proposing a realizable stable cosmic time crystal, although giving a different interpretation to the ``moving ground state", in which there is no motion because the system loses degrees of freedom. Furthermore, the model is free of pathologies such as propagating superluminal perturbations, negative energies, and perturbation instabilities.

\end{abstract}

\maketitle

\section{Introduction}  

Type IA Supernovae observations indicate that the Universe is experiencing an accelerated expansion \cite{RFC1}.
Furthermore, if one assumes that the Universe had a beginning, an early accelerated expansion phase called inflation has been postulated to overcome the horizon and flatness problems. Inflation has also been found to yield sensible initial conditions for the primordial inhomogeneous cosmological perturbations that gave rise to the present structures in the Universe.\footnote{In non-singular cosmological scenarios like bouncing models, these problems are not necessarily present, and the existence of an early inflationary phase might be unnecessary \cite{N1}.} Both in the early inflationary and the late accelerated phases, the Universe is approximately a de Sitter space-time, although the two phases are separated by a huge difference in energy density. In order to account for these scenarios, unusual actions --mostly inspired in possible UV completions-- involving scalar fields have been proposed, extending the bestiary of fundamental fields and often reproducing one expansion phase or the other, but not both. Thus, originally motivated by string theory, scalar fields with non canonical kinetic terms --\kit-essence fields--, which approximately describe the (A)dS geometry as an attractor solution, were introduced first to model inflation \cite{ADM1}, and dark energy later on \cite{COY1,ADS1}.

Nevertheless, the perks of a non-linear dynamic are shadowed by apparent unphysical features of the solutions: superluminal propagation of perturbations \cite{CCR1} (even though causality is still preserved \cite{BMV1}); negative energy and perturbative instabilities as the system evolves into regions where the Null Energy Condition (NEC) is violated \cite{LN1,SMM1}; and the presence of singularities in the dynamical motion \cite{JMW1}, leading to loss of hyperbolicity \cite{BLL1}, horizon formation \cite{GB1}, and caustics in wave propagation \cite{FKS02,Babichev16}.

Besides, \kit-essence provides an intriguing solution: a cosmological realization of Shapere and Wilczek's classical time crystal \cite{Shapere:2012nq}, whose ground state has broken time translation symmetry, only possible in systems with a multi-valued Hamiltonian. Time crystals have been actively investigated in condensed matter physics, albeit departing from the original proposal \cite{HS1}. On the other hand, cosmological versions were initially proposed in Ref. \cite{BHW1}, where the periodic behavior of the field would lead to new cosmological phases. However, in order to reach this ground state, the system must violate the NEC, the speed of sound squared becomes negative, and perturbations would grow exponentially \cite{EV1}. An effective field theory approach was able to regularize the system \cite{EM1}, at the cost of new free time-dependent parameters. Cosmic time crystals also emerged in a purely geometric Universe with noncommutative geometry corrections \cite{DPGP1}.

In this paper, we present a stable model that reaches the so called ``moving ground state" without requiring extra dynamical additions. We interpret the physical system not as a time crystal, but as a degeneracy in the dynamical structure of the model \cite{STZ1,ANZ1}. This greatly enriches the \kit--essence models, presenting the degeneracy as a new process in which the field that produces an early slow roll inflation evolves into a form of dark energy today (not an attractor solution anymore). Hence we have one field for both phases of accelerated expansion. Different proposal for that exist, for example using modified $f(R)$ gravity models  \cite{Nojiri2003,Cognola2007,Nojiri2007}, and phantom scalar fields \cite{Nojiri2005}.

Moreover, some other issues are alleviated, as the singularity in the motion is no longer a problem, but a dynamical feature that limits the motion in phase space to a bounded sector in such a way that the regions where the field is unstable and superluminal propagation occur, are inaccessible.

This alternative interpretation was explored in \cite{ANZ1}, where systems possessing a multi-valued Hamiltonian lose degrees of freedom as they get trapped on a surface of phase space where the dynamical equations degenerate. This happens because the single-valued branches of the momenta $p_i(q^j, \dot{q}^j)$ are separated by {\it{degeneracy surfaces}}, where the Hessian determinant $|\partial p_i/\partial \dot{q}^j|$ has a simple zero. From the equations of motion,
\begin{equation*}
    \frac{\partial p_i}{\partial \dot{q}^j}\, \ddot{q}^j = -\frac{\partial p_i}{\partial q^j} \, \dot{q}^j + \frac{\partial L}{\partial q^i},
\end{equation*}
it is clear that if the right side is nonzero, the system is subjected to an infinite acceleration that changes sign, therefore, being attracted or repelled by such surfaces, sticking to it in the first case. On the degeneracy surface, some degrees of freedom become gauge modes, and we interpret the ground state motion taking place at the surface simply as a gauge transformation in a system that is stuck there. This phenomenon of freezing out degrees of freedom was first observed in Lovelock gravity theories for $D>4$ \cite{Teitelboim:1987zz, Henneaux:1987zz}, and is a rather conspicuous feature of gravitation and supergravity theories where the dynamical loss of degrees of freedom represents a dynamical mechanism for dimensional reduction \cite{Hassaine:2003vq, Hassaine:2004pp, Miskovic:2005di, Zanelli:2005sa}

In Sec. \ref{sec2}, we present our model consisting of a FLRW Universe filled with a \kit--essence field. We review the requirements to produce an acceptable model, and how those requirements can be fulfilled by our proposal. In Sec. \ref{sec3}, the modified slow roll inflation is presented, together with the constraints in the parameters imposed by the fact that the field describes dark energy today. Then, in Sec. \ref{sec4}, the degeneracy mechanism is discussed, showing how the system loses degrees of freedom as it degenerates, producing a cosmological constant in the late universe. In Sec. \ref{sec5}, we examine the perturbations checking that they remain well behaved, as expected for a reasonable \kit--essence theory, to ensure stability once the field degenerates. We end up summarizing our conclusions with some remarks for future developments.

\section{\textit{K}-essence Cosmology}  
\label{sec2}

Consider a spatially flat FLRW space-time with metric $ds^2=-N(t)dt^2+a(t)(dx^2+dy^2+dz^2)$ filled with a \textit{k}-essence scalar field, where the lapse function $N(t)$ will be later set equal to one. The minisuperspace action can be written as
\begin{equation}
    S=\int\rmd t a^3\kappa\left[-\frac{1}{2N}\left(\frac{\dot{a}}{a}\right)^2 +N\kappa^2 \mathcal{L}\right].
\end{equation}
with $\kappa=\sqrt{4\pi G/3} = \sqrt{4\pi/3}/m_{\rm Pl}$ and $m_{\rm Pl}$ is the Planck mass. The matter Lagrangian density is given by
\begin{equation} \label{L2}
    \mathcal{L}=f(\phi)k(X)-V(\phi)\,,
\end{equation}
where the kinetic term is an \textit{a priori} arbitrary function $k(X)$ of the canonical kinetic term $X=(\nabla \phi)^2/2=\dot{\phi}^2/2N^2$, to be conveniently chosen for different purposes \cite{COY1,ADS1}. Moreover, we assume $f(\phi)>0$. Writing the energy-momentum tensor in analogy with a perfect fluid
\begin{equation}
    T_{\mu\nu}=(\varepsilon+p)u_{\mu}u_{\nu}-pg_{\mu\nu},
\end{equation}
the field's four-velocity being $u_{\mu}=\textrm{sgn}(\dot{\phi})\nabla_{\mu}\phi/\sqrt{2X}$. Then, it has pressure $p=\mathcal{L}$, and energy density
\begin{equation} \label{epsilon}
    \varepsilon=2X\mathcal{L}_{,X}-\mathcal{L}=f(2Xk_{,X}-k)+V.
\end{equation}
The dynamics is found by varying the action with respect to $\phi$ and $N$. Hereafter we choose the time coordinate so that $N=1$,
\begin{equation}
    H^2=\left(\frac{\dot{a}}{a}\right)^2=2\kappa^2\varepsilon,\quad \varepsilon,_{X}\ddot{\phi}=-3Hp,_{X}\dot{\phi}-\varepsilon,_{\phi},
\end{equation}
from which we see that the energy density is non negative $\varepsilon\geq 0$, and the degeneracy surface of $\phi$ is reached when $\varepsilon,_{X}=2X\mathcal{L},_{XX}+\mathcal{L},_{X}=f(2Xk,_{XX}+k,_{X})=0$.

As mentioned above, the presence of such uncanny kinetic term could account for an accelerated expansion through an attractor solution with an equation of state $w=p/\varepsilon = -1$. From Eqs. (\ref{L2}) and (\ref{epsilon}), requiring
\begin{equation}
    w=\frac{fk-V}{f(2Xk,_{X}-k)+V} = -1,
\end{equation}
implies that, for some finite time, $Xk,_{X} = 0$.

Another important constraint is due to the stability of the model: the speed of sound must be real,
\begin{equation}
    c^2_s=\frac{p,_{X}}{\varepsilon,_{X}}=\frac{k,_{X}}{2Xk,_{XX} + k,_{X}}\geq0,
\end{equation}
lest the perturbations grow exponentially.

The problem now is this: to reach the condition $\varepsilon,_{X}=0$ passing through a configuration with $w=-1$, the solution must cross $k,_{X}=0$ with $k,_{XX}\neq0$ there, forcing $c^2_s$ to change sign, which produces a gradient instability. In order to avoid this, one could require $k(X)$ to be such that $c^2_s=\gamma$, a positive constant, which gives $k,_{X}\propto X^{\frac{1-\gamma}{2\gamma}}$. However, for $0<\gamma<1$, $\varepsilon,_{X}\propto k,_{X}/\gamma$ is null only for $X=\dot{\phi}^2=0$, which is not a simple zero, hence the acceleration does not change sign as it passes through this point and the solutions are not forced to end in the surface $\varepsilon,_{X}=0$, i.e., the system does not degenerate.

Another possibility is that both $p,_{X}$ and $\varepsilon,_{X}$ approach zero for some value $X=X_d\neq 0$, i.e.: $k,_{X}(X_d)=0$ and $k,_{XX}(X_d)=0$. Under these conditions, expanding the function $k(X)$ around $X_d$, we find
\begin{equation} \label{ansatz}
    \mathcal{L}=f(\phi)(X-X_d)^n-V(\phi),
\end{equation}
with $n\geq 3$, which can be an approximation of a more general theory near the degeneracy surface. From ansatz \eqref{ansatz}, one finds a stable model that degenerates and behaves as an Universe dominated by a cosmological constant, fitting the description of today's accelerated expansion thereafter. 

In order to simplify the treatment and track the system dependence on the parameters, we set $n=3$ and rewrite the parameters as powers of mass dimension quantities: $f=\mu_f^{-8}$, and $X_d=\mu_d^4$. We also rescale time and the scalar field into dimensionless quantities through $\rmd\Bar{t}=\mu_d^4\kappa^3\rmd t$ and $\Bar{\phi}=\mu_d^2\kappa^3\phi$. Then, the action takes the form
\begin{multline}
    S=\int\rmd\Bar{t}a^3\left[-\frac{\lambda^4_1}{2N}\left(\frac{1}{a}\frac{\rmd a}{\rmd \Bar{t}}\right)^2\right.\\ +\lambda_2N(\Bar{X}-1)^3-NV(\Bar{\phi}) \Bigg],
\end{multline}
where $\lambda_1=\mu_d\kappa$, and $\lambda_2=(\mu_d/\mu_f)^{8}$. Moreover, we have incorporated a $\mu_d^{-4}$ term in the potential, so that it becomes dimensionless. The parameters in the dimensionless potential $\Bar{V}=\Bar{\alpha}\Bar{\phi}^D$, will be associated with the dimensional ones, $V=\alpha\phi^D$, through $\Bar{\alpha}=\alpha/(\mu_d^{4+2D}\kappa^{3D})$. The relation between the dimensional and dimensionless energy density and pressure is $\varepsilon=\mu^4_d\Bar{\varepsilon}$ and $p=\mu^4_d\Bar{p}$.

The quantities of interest and the equations of motion are (from now on we remove the bars for simplicity)
\begin{equation}
    p=\mathcal{L}=\lambda_2\Delta^3-V,\hspace{3mm}\varepsilon=\lambda_2\Delta^2(5X+1)+V,
    \label{eq7}
\end{equation}
\begin{equation}
    H^2=\frac{2}{\lambda_1^4}\varepsilon
    \label{eq10}
\end{equation}
\begin{align}
    3\lambda_2\Delta(5X-1)\ddot{\phi}&=-9H\lambda_2\Delta^2\dot{\phi}\nonumber
    \\&-\frac{d\lambda_2}{d\phi}\Delta^2(5X+1)-V_{,\phi}.
    \label{eq11}
\end{align}
The degeneracies occur at $\Delta=X-1=0$ and $X=1/5$. We see that for the system to fall into the degeneracy at $X=1$, it requires $V_{,\phi}(\phi_d)\neq0$, where $\phi_d$ is the value of $\phi$ for $X=1$, hence the potential is crucial for the degeneracy to occur. 

From now on we set $\lambda_2=const$ for simplicity. It is useful to write the dynamical equation as a first order differential equation
\begin{multline}
    \frac{\rmd X}{\rmd\phi}=-\textrm{sgn}(\dot{\phi})\frac{6\Delta}{5X-1}\sqrt{\frac{\lambda_2}{\lambda^4_1}X\left[\Delta^2(5X+1)+\frac{V}{\lambda_2}\right]}\\-\frac{V_{,\phi}}{3\lambda_2\Delta(5X-1)}.
    \label{eq12}
\end{multline}
Note that the Hubble drag term is proportional to $\Delta$ and to $\lambda_2$, and is negligible for $\Delta\approx0$. Therefore, increasing $\lambda_2$ increases the expansion effect on the field while suppressing the interaction terms coming from $V$. Also
\begin{equation}
    c^2_s=\frac{\Delta}{5X-1},
\end{equation}
vanishes for $X=1$, as required. Instabilities occur for $1/5<X<1$, and $c^2_s\rightarrow \pm \infty$ as $X$ approaches $1/5$ from above or from below. However, the degeneracy surfaces have the remarkable property to divide the phase space of the system into disjoint regions \cite{STZ1,ANZ1} -- then, in our model, stable systems remain stable. Therefore, we are interested in the region where $X>1$, reaching $\Delta=0$ with positive $c_s^2$, where the system is healthy. One can also see that in this region $c^2_s<1$ and hence there is no superluminal propagation.

Finally, the presence of a potential is not only important for the system to degenerate, but also to lead to an attractor solution, which appears when the RHS of Eq.~\eqref{eq12} is null for $\Delta \neq 0$, and the potential term dominates over the kinetic one, yielding an early accelerated expansion phase with a much higher energy density.

\section{The early Universe}  
\label{sec3}

The field loses all its degrees of freedom at the degenerate surface $\Delta=0$, converting $V(\phi)$ into an effective cosmological constant, since from Eq.~\eqref{eq7} one gets $p=-\varepsilon$ for any potential $V$. Therefore, this stage can be very useful to explain the accelerated expansion today, but not an early one: an inflationary phase of the Universe must have a finite duration for matter to cluster into the structures we observe.

Nevertheless, the non-canonical kinetic term in our model can also lead to the desired behavior of a late de Sitter phase through a different physical process, allowing for a concomitant slow roll inflation. Hence, the field $\phi$ can, by itself, describe the two phases of accelerated expansion of the Universe.

Before describing the mechanism itself, following \cite{M1} let us consider the possible initial conditions of the model -- how the Universe began. If we assume something like an initial singularity, $\varepsilon\rightarrow\infty$ and $X\rightarrow\infty$, we get
\begin{equation}
    \frac{\rmd\dot{\phi}}{\rmd\phi}\approx-\sqrt{\frac{3\lambda_2}{40\lambda^4_1}}\dot{\phi}^3
\end{equation}
Integrating, and assuming an initial value $\dot{\phi}_i\rightarrow\infty$, we find that $\dot{\phi}\propto(\phi-\phi_i)^{-1/2}$. As the time derivative of the field decays faster than its value, the field falls into the attractor with a small deviation from its initial value, enlarging the possible set of initial conditions.

Remarkably, our model also allows for a very interesting beginning. In the degeneracy surface $\Delta = 0$, as the field becomes a cosmological constant, the cosmological solution is a de Sitter universe. Hence, near the repulsive part of the degeneracy surface the Universe shall emerge from a singularity free de Sitter space. In the neighborhood of the degeneracy surface, Eq.~\eqref{eq12} yields
\begin{equation}
    \frac{\rmd X}{\rmd \phi}\approx-\frac{V_{,\phi}}{3\lambda_2\Delta(5X-1)}.
\end{equation}
Again, the kinetic term grows fast, until it reaches the attractor solution, whilst the value of the field does not change much ($V_{,\phi}$ and $\dot{\phi}$ must have opposite signs for the surface to be repulsive, so that $\dot{\phi}$ and $\rmd X/\rmd\phi$ must have the same sign).

Assume the attractor solution to be in the region $X>1$, and that it allows 
\begin{equation}
    V\gg \lambda_2\Delta^2(5X+1),
    \label{eq121}
\end{equation}
which implies
\begin{equation}
    V\gg \lambda_2\Delta^3.
    \label{eq122}
\end{equation}
Then, reaching the attractor solution either from a singularity or from a de Sitter space, space-time will start to inflate as in the slow roll scenario, $p\approx-\varepsilon$, see Eqs.~(\ref{eq121},\ref{eq122},\ref{eq7}), with $\phi_i$ almost unchanged.

Inflation shall last for some time with $\ddot{\phi}\approx0$. The above conditions on the dynamical equation for the field (\ref{eq10},\ref{eq11}) yields
\begin{equation}
9\lambda_2\Delta^2\dot{\phi}\sqrt{\frac{2}{\lambda^4_1}V}\approx -V_{,\phi}.
\label{eq125}   
\end{equation}
Squaring this, one gets
\begin{equation}
    \Delta^4X\approx\frac{\lambda^4_1}{324\lambda_2}\frac{V^2_{,\phi}}{V},
    \label{eq14}
\end{equation}
which has a solution for $X>1$ with $V>0$. This is because $h(X)\equiv\Delta^4X-\lambda^4_1 V^2_{,\phi}/(324\lambda_2 V)$ has a local minimum at $X=1$, increasing monotonically afterwards. Hence, as $h(1)<0$, $h(X)=0$ for some $X>1$.

Furthermore, as $\varepsilon\approx V$, and using Eqs.~(\ref{eq10},\ref{eq125}), one gets
\begin{equation}
    H=\dot{\phi}\frac{\rmd\textrm{ln}\,a}{\rmd\phi}\approx-\frac{V_{,\phi}}{9\lambda_2\Delta^2\sqrt{2V/\lambda^4_1}}\frac{\rmd\textrm{ln}\,a}{\rmd\phi}\approx\sqrt{\frac{2}{\lambda^4_1}V}
\end{equation}
which, as $\ddot{\phi}=dX/d\phi\approx0$ implies $X\approx cte$, can be easily integrated to give the scale factor
\begin{equation}
    a(t)\approx a_i\textrm{exp}\left(\frac{18\lambda_2 \Delta^2}{\lambda^4_1}\int^{\phi_i}_\phi \frac{V}{V_{,\phi}}\rmd\phi\right),
    \label{eq20}
\end{equation}
which has to grow more than 75 e-folds \cite{M1} in order for the mechanism to be realistic.

The situation described here is very general: given a potential $V$ that allows a nearly flat attractor solution, $X\approx const$, and for large values of $|\phi|$ so that condition (\ref{eq121}) is satisfied, then a large set of initial conditions produce inflation. Moreover, as one can see from the equations of motion (\ref{eq12}), after inflation ends $X$ decays and the system inevitably falls into the degenerate surface $\Delta=0$, where the scalar field freezes, turning $V(\phi)$ into a cosmological constant.

Another relevant remark is that all phase space orbits that reach the inflationary solution pass through approximately the same points, falling into the degeneracy surface with the same value for the energy density $\mu^4_d\varepsilon_d$, corresponding to the present value of the cosmological constant. This is crucial as it does not require finely tuned initial conditions and the same parameters are responsible for both phases of accelerated expansion.
%
\subsection{An example: $V(\phi)=\mu^2 \phi^2/2$.} 

The above assertions become more comprehensible with the simplest dimensionless potential $V=\mu^2\phi^2/2$, with which eq (\ref{eq12}) takes the form
\begin{multline}
    \frac{\rmd X}{\rmd\phi}=-\alpha_1\left\{\textrm{sgn}(\dot{\phi})\frac{6\Delta}{5X-1}\sqrt{X\frac{\alpha_2}{\alpha_1}\left[\frac{\Delta^2(5X+1)}{\alpha_1}+\frac{\phi^2}{2}\right]}\right.\\\left.+\frac{\phi}{3\Delta(5X-1)}\right\}\,,
    \label{eq21}
\end{multline}
where the parameters are now
\begin{equation}
    \alpha_1=\frac{\mu^2}{\lambda_2}=\frac{\mu_f^8m^2}{\mu_d^{16}\kappa^6},\hspace{5mm}\alpha_2=\frac{\lambda_2}{\lambda^4_1}=\frac{\mu_d^4}{\mu_f^8\kappa^4}.
\end{equation}
The real mass of the field is given by the dimensionless parameter $\mu=m/(\mu_d^4\kappa^3)$. Eq. (\ref{eq14}) now becomes
\begin{equation}
    \Delta^4X\approx \frac{\alpha_1}{162\alpha_2},
    \label{eq23}
\end{equation}
which is a fifth degree polynomial that cannot be algebraically solved for $X$. However, since it does not depend explicitly on $\phi$, it has a constant solution $X=X_{at}$, which grows with the parameters as $[\alpha_1/(162\alpha_2)]^{1/5}$.
Inflation ends when the potential becomes of the order of the kinetic term. However, when compared with its very large initial value, the field will be negligible in the final state so, we assume that it ends with $\phi_f\approx0$. Then, from eq. (\ref{eq20}) we find the number of e-folds

%
%
%
\begin{equation}
    \textrm{N}=\textrm{ln}\frac{a_f}{a_i}=\sqrt{\frac{2}{X_{at}}}\frac{\mu\phi^2_i}{\lambda^2_1} .
\end{equation}
The minimum necessary, $N>75$, can be achieved by choosing a large enough $\phi_i$, bounded by the requirement that the energy density remain subplanckian: $\phi^2_i\lessapprox10\mu^4_d\kappa^2 /m^2$. Such requirements constraint $X_{at}\lessapprox10^{-1}/(m^2\mu^4_d\kappa^6)$, which can be easily satisfied.

As stated, all inflationary solutions degenerate at the same point with dimensional energy density $\mu_d^4\varepsilon_d=\mu_d^4\mu^2\phi_d^2/2=m^2\phi_d^2/(\mu^8_d\kappa^6)$, which is fixed by the free parameters of the model. Therefore, to find this dependence, we shall calculate the degenerate field $\phi_d$ for the inflationary solutions, which cannot be done analytically in the general case, however, can be done through some approximations.

\begin{figure}[!ht]
\includegraphics[width=0.5\textwidth]{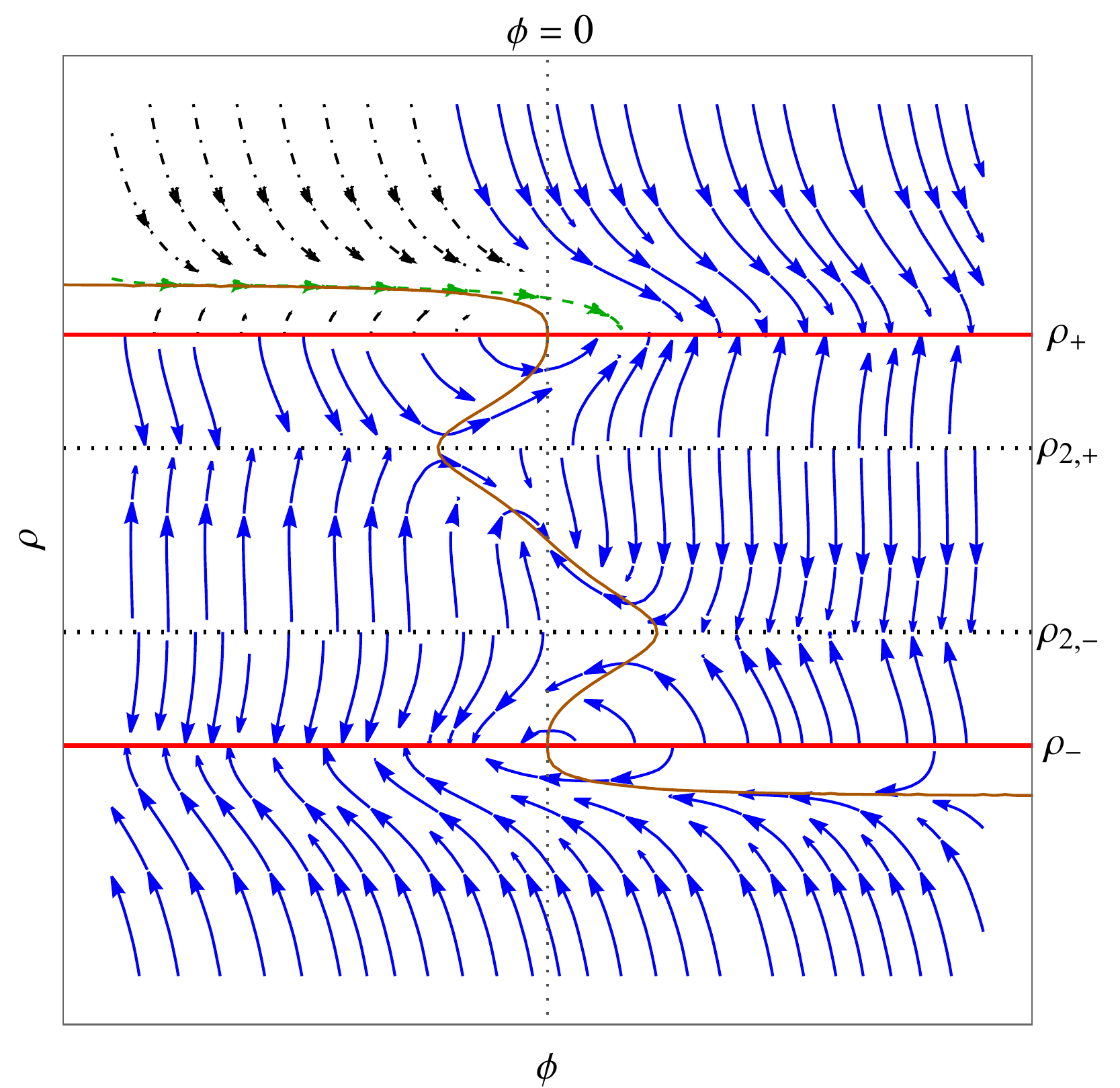}
\caption{Stream plot of the non-canonical phase space, where we identify $\rho=\dot{\phi}$, see sec. \ref{sec4}, for constant $\lambda_2$ and $V=\mu^2\phi^2/2$. The red full line and the black dotted line are the degenerate surfaces with $\rho=\rho_{\pm}=\pm\sqrt{2}$ and $\rho=\rho_{2,\pm}=\pm\sqrt{2/5}$, respectively. The dashed green stream line is the attractor inflationary solution, while the black dot--dashed streams are the ones that fall into it, coming from an initial singularity or from the repulsive part of the degeneracy surface. The full orange line denotes the curve $\rmd X/\rmd\phi=0$, which is not a solution of the dynamical equation.}
\label{fig1}
\end{figure}

For the approximations, notice that $h_0(X)=X(X-1)^4$, the function on the LHS of (\ref{eq23}), crosses the horizontal axis with slope 1 at $X=0$, it has a local maximum at $X=1/5$ ($h_0(1/5)\approx 0,08$), and is tangent to the axis at $X=1$. It is bounded by two functions: grows as $(X-1)^4$ for $1<X<2$ and as $X^5$ for $X\gg 2$. Thence, we will use these two limiting behaviors as approximations. First, $h_0\approx X_{at}^5$ when $\alpha_1\gg10^6\alpha_2$, because then $X_{at}=(\alpha_1/162 \alpha_2)^{1/5}\gg2$. Integrating $X$ from $X_{at}$ to $X=1$, one can notice that the term under the square root becomes negligible, as it is multiplied by $\alpha_2/\alpha_1 \approx 10^{-6}$. We can now integrate, assuming that $X=X_{at}$ for $\phi=0$,
\begin{equation}
    \frac{5X^3_{at}}{3}-3X^2_{at}+X_{at}-\frac{1}{3}\approx\frac{5X^3_{at}}{3}=\frac{\alpha_1}{6}\phi^2_d,
\end{equation}
with that we find $\phi_d\propto(\alpha^2_1\alpha^3_2)^{-1/10}$.

Whereas, if $\alpha_1\ll10^2\alpha_2$, then $h_0(X)\approx(X-1)^4=x^4\ll 1$ (For a very small value of the RHS of (25), $h_0(X)=\delta$, there are 3 values of $X$: $X\approx \delta$, $X \approx 1\pm \delta^{1/4}$, but only one for $X>1$). Now integrate from $x_{at}$ to $x=0$. When $x=x_{at}$, the first term inside the square brackets on the right hand side of eq. (\ref{eq21}) will be approximately of order one, while the second term in square brackets and the last therm on RHS will go as $(\alpha_1/\alpha_2)^{-1/4}\gg1$. Notwithstanding, when $x$ gets smaller, the former goes as $x$, being dominated by the latter that goes as $x^{-1}$ and, as we are interested just in the parameter dependence, only the last term survives. Integrating from $(0,x_{at})$ to $(\phi_d,0)$ 
\begin{equation}
    \alpha_1\phi^2_d=12x^2_{at},
\end{equation}
so that, in this case $\phi_d\propto(\alpha_1\alpha_2)^{-1/4}$.

As the degenerate configuration of the system is responsible for the present accelerated expansion, its energy density should be the same of the cosmological constant $\mu_d^4\varepsilon_d/m_{pl}^4=m^2\phi_d^2/(m_{pl}^4\mu^8_d\kappa^6)=\varepsilon_{\Lambda}/m_{pl}^4\approx10^{-123}$, constraining the values of our parameters:

\begin{itemize}
    \item $\alpha_1\gg10^6\alpha_2$ means
\begin{equation}
    \left(\frac{m}{m_{pl}}\right)^2\left(\frac{\mu_f}{m_{pl}}\right)^{16}\left(\frac{m_{pl}}{\mu_d}\right)^{20}\gg 10^6,
    \label{eq29}
\end{equation}
and the requirement of the energy density to be that of the cosmological constant
\begin{equation}
    \left(\frac{m}{m_{pl}}\right)^{6/5}\left(\frac{\mu_f}{m_{pl}}\right)^{8/5}\approx10^{-123}.
\end{equation}

\item While, when $\alpha_1\ll10^2\alpha_2$,
\begin{equation}
    \left(\frac{m}{m_{pl}}\right)^2\left(\frac{\mu_f}{m_{pl}}\right)^{16}\left(\frac{m_{pl}}{\mu_d}\right)^{20}\ll 10^2,
\end{equation}
and
\begin{equation}
    \left(\frac{m}{m_{pl}}\right)\left(\frac{\mu_d}{m_{pl}}\right)^{2}\approx10^{-123}.
\end{equation}
\end{itemize}

Note that the admissible orders of magnitude for the free parameters are not sensitive to the approximations that have been use. They indicate a very small scalar field mass, e.g. $m=10^{-22}\textrm{eV}=10^{-50}m_{pl}$, the mass of fuzzy dark matter models, yielding $\mu_d\approx10^{-37}m_{pl}$, and $\mu_f\approx10^{-40}m_{pl}$. It is also possible to have a larger mass, $m=10^{-16}\,m_{pl} \sim 1\,\textrm{TeV}$, which imposes much smaller values, $\mu_d=10^{-53}\,m_{pl}=10^{-25}\,\textrm{eV}$ and $\mu_f \approx10^{-65}\,m_{pl}$.

In any of these scenarios, both $\mu_d$ and $\mu_f$ have to be very small. One can infer easily the consequences of these small numbers in flat space-time. Going back to dimensional quantities:
\begin{equation}
    \ddot{\phi}=-\frac{\mu^8_f V_{,\phi}}{3(X-\mu^4_d)(5X-\mu^4_d)}.
\end{equation}
If $X\gg\mu^8_fV_{,\phi}$ and $X>\mu^4_d$ (both very small quantities), the field will have a constant velocity as the interaction will be dumped. Otherwise, if $X\approx\mu^4_d$, the field will be very near the degeneracy surface, being rapidly dragged into it or out of it, losing dynamics or rapidly reaching the high $X$ regime with a constant velocity, respectively.

Small values of the free parameters commonly appear in any attempt to accommodate the very small value of the cosmological constant in the Standard Cosmological Model, and they do not constitute a satisfactory explanation for it. However there is ample room for improvement in the model that will not change the overall idea of the degeneracy and can ease this problem. This matter will be addressed in the next subsection and in the final remarks.

\subsection{Another example: $V(\phi)=\beta^4 \phi^4/4-\mu^2\phi^2/2$.}

To test the generality of the model, we introduce a Higgs type potential $V(\phi)=\beta^4 \phi^4/4-\mu^2\phi^2/2$, in which\footnote{From now on, in this section, we assume that $\dot{\phi}>0$, remembering that the solutions are the same if $(\phi,\dot{\phi})\rightarrow(-\phi,-\dot{\phi})$.}

\begin{multline}
    \frac{\rmd X}{\rmd\phi}=-\frac{6\Delta}{5X-1}\sqrt{\frac{\lambda_2}{\lambda^4_1}X\left[\Delta^2(5X+1)+\frac{\beta^4\phi^4}{4\lambda_2}-\frac{\mu^2\phi^2}{2\lambda_2}\right]}\\-\frac{\beta^4\phi^3-\mu^2\phi}{3\lambda_2\Delta(5X-1)}.
    \label{eq37}
\end{multline}
Now the dimensionful parameters in the potential are related by $\beta=\omega/(\mu^3_d\kappa^3)$, and $\mu=m/(\mu^4_d\kappa^3)$ as before. Inflation occurs for large values of the potential at large $|\phi|$. In this regime, we can assume $V\approx\beta^4\phi^4/4$ and from (\ref{eq14}),
\begin{equation}
    81\Delta^4X\approx\frac{\lambda^4_1}{\lambda_2}\beta^4\phi^2.
    \label{eq38}
\end{equation}
Then $X$ will not be a constant--a finite $\dot{\phi}$ implies that $\phi$ and, consequently, the above equation changes over time. However, when $|\phi|$ is large, it will be approximately constant, which can be seen by noting that
\begin{equation}
    \frac{\rmd X}{\rmd\phi}\approx-\left(\frac{3}{\lambda^2_1}\sqrt{X}\Delta+\frac{\beta^2\phi}{3\lambda_2\Delta}\right)\frac{\beta^2\phi^2}{5X-1}.
\end{equation}
If the term in parenthesis is nonzero, $|\rmd X/\rmd\phi|\gg1$, and $X$ will decrease (or grow, if it starts from the degenerate surface, remember that we assume $\Delta>1$) until the term in parenthesis reaches zero at $\phi=-9\lambda_2\Delta^2\sqrt{X}/\beta^2$ and $\rmd X/\rmd\phi\approx0$ (remember that $\phi$ and $\dot{\phi}$ have opposite signs in the inflationary phase). Moreover, we can assume the RHS of (\ref{eq38}) to be much larger than one, so that $X^5_{at}\approx\lambda^4_1\beta^4\phi^2/(81\lambda_2)$.
\begin{figure}[!ht]
\includegraphics[width=0.5\textwidth]{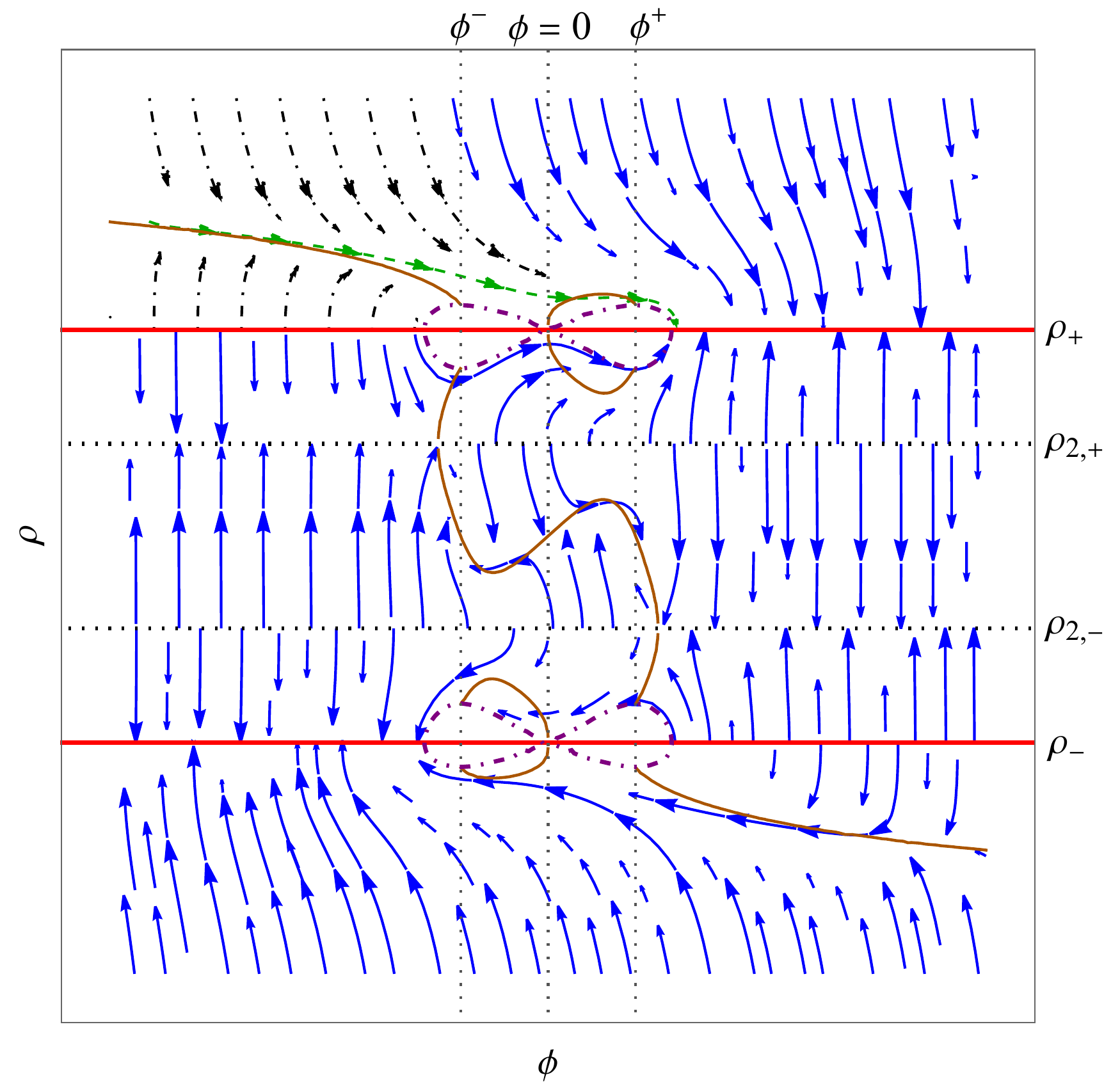}
\caption{Stream plot of the non-canonical phase space, where we identify $\rho=\dot{\phi}$, see sec. \ref{sec4}, for $\lambda_2=cte$ and $V=\beta^4\phi^4/4-\mu^2\phi^2/2$. The red full line and the black dotted line are the degenerate surfaces with $\rho=\rho_{\pm}=\pm\sqrt{2}$ and $\rho=\rho_{2,\pm}=\pm\sqrt{2/5}$, respectively. The dashed green stream line is the attractor inflationary solution, while the black dot--dashed streams are the ones that fall into it, coming from an initial singularity or from the repulsive part of the degeneracy surface. Again, the full orange line denotes the curve $\rmd X/\rmd\phi=0$, which is not a solution of the dynamical equation and here the dot-dashed purple curve denotes $\varepsilon=0$, whose interior contains no orbits.}
\label{fig2}
\end{figure}

Inflation ends when condition (\ref{eq121}) is not satisfied anymore, which implies
\begin{equation}
    \frac{\beta^4}{4}\phi^4_f\approx\lambda_2\Delta^2(5X+1)\approx5\lambda_2X_{at}^3,
\end{equation}
hence $\phi_f\approx\pm\left(\lambda_2\lambda^6_1/\beta^4\right)^{1/7}$. With that we can calculate the number of e-folds from Eq. (\ref{eq20})
\begin{equation}
    N=\frac{9\lambda_2}{2\lambda^4_1}X^2_{at}\left[\phi^2_i-\left(\frac{\lambda_2\lambda^6_1}{\beta^4}\right)^{2/7}\right]\approx\frac{9^{1/5}}{4}\left(\frac{\lambda^5_2\beta^8}{\lambda_1^{12}}\right)^{1/7}\phi^2_i,
\end{equation}
which, again, must be higher than 75, requiring a large enough value for $\phi^2_i$.

Let us now estimate the value of the final cosmological constant for orbits that experience sufficient inflation when they reach the degeneracy surface. We will evaluate it analytically using the information presented in the stream plot for the Higgs-like potential shown in FIG.~\ref{fig2}. Note that this figure exhibits an attractor solution (the dashed green curve) where inflation occurs for $\phi$ large and negative. This attractor is approximately the curve $\rmd X/\rmd\phi=0$ (full orange line), and that it reaches the degeneracy surface very near the curve satisfying $\varepsilon=0$ (the purple dot-dashed curve). However, after the solution departs from the orange curve we lose its track, as the previous approximations does not hold and the motion enters a non-linear regime. Near the degeneracy surface, the equations of motion could be integrated approximately if we identify the orbits that reach the degeneracy coming from the attractor. 
The equation of motion in the coordinate $x=X-1$ is
\begin{equation}
   \frac{\rmd x}{\rmd\phi}= -\frac{6x}{\lambda^2_1(5x+4)}\sqrt{(x+1)\varepsilon}-\frac{V_{,\phi}}{3\lambda_2x(5x+4)},
    \label{eq42}
\end{equation}
and the energy density
\begin{equation}
   \varepsilon= \lambda_2x^2(5x+6)+V.
   \label{eq43}
\end{equation}

In order to find which orbits come from the attractor, we assume they are close to the curve $\varepsilon=0$ for $\phi>0$, and therefore the Hubble drag (the first term in the RHS of Eq.~(\ref{eq42})) is small, and is eventually overwhelmed by the last term, somewhere between $\phi=0$ and $\phi^{+}=\mu^2/\beta^2$, where $V_{,\phi}<0$ -- the dot-dashed green stream line meets again the orange curve between these values, see FIG.~\ref{fig2} -- and in this point $\rmd x/\rmd\phi=0$. Take this intersection is approximately at the midpoint $\phi_i=\phi^{+}/2$. Then, from Eq.~(\ref{eq42}) we get
\begin{equation}
    -18x^2_i\frac{\lambda_2}{\lambda^2_1}\sqrt{(x_i+1)\varepsilon_i}=V_{,\phi}(\phi_i).
    \label{eq44}
\end{equation}
As we are near the degeneracy, $x\ll 1$, and squaring Eq.~(\ref{eq44}) we get
\begin{equation}
    324(\lambda_2x^2_i)^2\varepsilon_i=\lambda^4_1V^2_{,\phi},
    \label{eq45}
\end{equation}
and from \eqref{eq43}, $\varepsilon_i \approx 6\lambda_2x_i^2+V$, so that
\begin{equation}
\lambda_2x_i^2 \approx \frac{1}{6}(\varepsilon_i-V).
    \label{eq46}
\end{equation}
Moreover, near the degeneracy also $\varepsilon\ll1$ and the Hubble drag, which is of order $\mathcal{O}(x\varepsilon^{1/2})$, can be neglected in the subsequent evolution of the system, which means that the energy will be approximately conserved until it degenerates: $\varepsilon_d=\varepsilon_i$. Substituting (\ref{eq46}) in (\ref{eq45}) and using $V(\phi_i)=-7\mu^4/(64\beta^4)$ and $V_{,\phi}(\phi_i)=-3\mu^3/(8\beta^2)$ we find an equation for the degenerate energy density:
\begin{equation}
    \left(\frac{\beta^4\varepsilon_d}{\mu^4}+\frac{7}{64}\right)^2\varepsilon_d=\frac{\lambda^4_1}{64}\frac{\beta^4}{\mu^2},
\end{equation}
which need more approximations to be solved. First, if $\beta^4\varepsilon_d/\mu^4\ll1$, then $\varepsilon_d\propto\lambda^4_1\beta^4/\mu^2$. Again, as in the previous subsection, we shall compare the dimensional energy density with the energy density of the cosmological constant driving the accelerated expansion today: $\mu^4_d\kappa^4\varepsilon_d=\varepsilon_{\Lambda}\kappa^4$. Substituting $\varepsilon_d$ found above together with the definitions of the dimensionless parameters the physical parameters must satisfy
\begin{align}
    \omega^4\left(\frac{\mu_d}{m_{Pl}}\right)^4\left(\frac{m_{Pl}}{m}\right)^2\approx10^{-123},\\
\end{align}
In this approximation a larger mass is more helpful: take $m=10^{-16}m_{Pl}=1\,\textrm{TeV}$, as in the previous subsection. Choosing also that $\omega^4\approx10^{-1}$, as required in the Higgs potential for it to be renormalizable, we find that $\mu_d\approx10^{-38}m_{Pl}$ for the field to be responsible for today's accelerated expansion, alleviating somewhat the smallness of the parameter.
Now, for such values to satisfy the condition $\beta^4\varepsilon_d/\mu^4\ll1$ we shall have
\begin{equation}
    \omega^4\left(\frac{m_{Pl}}{m}\right)^4\left(\frac{\varepsilon_{\Lambda}}{m^4_{Pl}}\right)\approx10^{-123}\omega^4\left(\frac{m_{Pl}}{m}\right)^4\ll 1,
\end{equation}
which will be satisfied.
Another option would be $\beta^4\varepsilon_d/\mu^4\gg1$, so that $\varepsilon_d^3\approx \lambda^4_1\mu^6/(64\beta^4)$. Then, again taking the dimensional energy density given by the present cosmological constant, one finds
\begin{equation}
    \frac{1}{\omega^{\frac{4}{3}}}\left(\frac{\mu_d}{m_{Pl}}\right)^\frac{4}{3}\left(\frac{m}{m_{Pl}}\right)^2\approx10^{-123}.
\end{equation}
For an estimate, take the mass to be again the one for fuzzy dark matter models $m=10^{-50}m_{Pl}$, then
\begin{equation}
\frac{1}{\omega}\left(\frac{\mu_d}{m_{Pl}}\right)\approx10^{-17}.
\end{equation}
If again $\omega=10^{-1/4}\approx1$, then $\mu_d=10^{-17}m_{Pl}=10^2$\,GeV, which is significantly greater than the one found in the previous subsection, while $\mu_f$ is not constrained by the late behavior of the field.
Moreover $\beta^4\varepsilon_d/\mu^4\gg1$ means, for the dimensional parameters,
\begin{equation}
     \left(\frac{m}{m_{Pl}}\right)^2\left(\frac{m^4_{Pl}}{\varepsilon_{\Lambda}}\right)^2\approx10^{246}\left(\frac{m}{m_{Pl}}\right)^2\gg 1,
\end{equation}
being again easily satisfied.

The Higgs-type potential alleviates the smallness of the parameter $\mu_d$, which defines the scale of the degenerate surface, while not constraining $\mu_f$ at the cost of the new parameter $\omega$. As said above, further developments needed to bring the model closer to the real universe could help resolve this problem, as will be discussed in the final remarks.
%

\section{The late Universe}  
\label{sec4}

Different interpretations are given to the singularities appearing in the dynamics of k-essence fields and similar systems: Terminating singularities, caustics, sonic horizons, cosmic time crystals. Our proposal to consider such systems as degenerate provides fruitful new insights with a different perspective from the standard picture: it does not reach a moving ground state, nor is it ill-defined; the system just freezes out, losing degrees of freedom in a dynamical dimensional reduction process \cite{STZ1,ANZ1,Hassaine:2003vq, Hassaine:2004pp, Miskovic:2005di, Zanelli:2005sa}. Hence, it is through this loss of degrees of freedom that the k-essence field turns into the cosmological constant that drives the accelerated expansion today. 

As the canonical phase space is apparently ill-defined, a first order Lagrangian $L=p_{\phi}(\phi,\rho)\dot{\phi}-H(\phi,\rho)$ in the phase space spanned by $(\phi,\rho)$ is useful, where $\rho:=\dot{\phi}$,
\begin{align} \label{L1}
    L=a^3\left(\lambda_2 k_{,X}\rho\dot{\phi}-\varepsilon\right),
\end{align}
$X=\rho^2/2$ throughout this section. The dynamical equations are equivalent to (\ref{eq7}-\ref{eq11}),
\begin{align}
    \varepsilon_{,X}\dot{\rho} + 3Hp_{,X}\rho + \varepsilon_{,\phi}&=0,\\
    \varepsilon_{,X}(\dot{\phi}-\rho)&=0,
\end{align}
where the second equation identifies $\rho$ as $\dot{\phi}$. 
Within this formalism the character of the degeneracy surface $\varepsilon_{,X}=3\lambda_2\Delta(5X-1)=0$ is given by the sign of $\Phi=j^in_i$, where $\Vec{j}=\varepsilon_{,X}(\dot{\phi},\dot{\rho})$ the Liouville current and $n_i=\partial_i\varepsilon_{,X}$ the normal to the surface. The surface is repulsive or attractive depending on whether the flux $\Phi$ is positive or negative, respectively. In this case,
\begin{equation}
    \Phi=a^6\rho\,\varepsilon_{,XX}\left(\varepsilon_{,\phi}+3Hp_{,X}\rho\right)\vline_{(\phi,\rho)\rightarrow(\phi_d,\rho_d)},
\end{equation}
where $(\phi_d,\rho_d)$ is the point where the orbit intersects the degenerate surface \cite{STZ1,ANZ1}.

For $\Delta=0$ (so that $\rho_{d}=\rho_{\pm}=\pm\sqrt{2}$) it simplifies to
\begin{equation}
    \Phi=-12a^6V_{,\phi}(\phi_d)\rho_{\pm}.
\end{equation}
Given $\rho_{\pm}$, the flux depends only on the sign of $V_{,\phi}(\phi_d)$. Choosing $V=\mu^2\phi^2/2$, the degeneracy surface $\rho_+$ will be repulsive for negative values of the field, changing character when $\phi=0$, becoming attractive (the opposite for $\rho_-$). In addition, notice that the Hubble drag term is negligible near the surface, as it is proportional to $\Delta^2$.
For a given potential, a clear dynamical picture is drawn by the stream plot in the non-canonical phase space $(\phi,\rho)$. Figure 1  describes the evolution of the system for the harmonic potential and Figure \ref{fig2} does the same for the Higgs-type one. The overall behavior --generic inflationary evolution and final degenerate state for the current accelerated expansion-- does not change much for different choices of the potential. The system can come from an initial singularity, $X\rightarrow\infty$, or from the repulsive part of the degenerate surface, $X=1$, both with an arbitrary value of $\phi_i$ -- given that it falls and stays into the attractor the necessary amount of time. There, $X=X_{at}\approx const.$ and $|\phi|$ decreases until inflation is over and the field leaves the attractor. After that the system reaches the degeneracy surface, $\phi$ freezes on a fixed value which depends on the free parameters of the theory, the equation of state is $w=-1$ and the energy density $\mu^4_d\varepsilon_d=\varepsilon_\Lambda$. In the meantime, between the end of inflation and the degeneracy, radiation, --and possibly matter as well (if the field does not degenerate first)--, will begin to dominate. The only change in the dynamics of the field is through $H$, which does not contribute significantly near the degeneracy. The degenerate value of the scalar field determines the cosmological constant. If this value is small, the degenerate field will quietly wait while cosmological perturbations grow and structures form, for its time to dominate again the Universe evolution, and guide another accelerated expanding phase.
Notice that the system is symmetric under $(\phi,\rho)\rightarrow(-\phi,-\rho)$, so that orbits coming from the upper left are the same as the ones from the bottom right (the plot is invariant under rotations by $\pi$).

Once in the attractive part of the degeneracy surface, the system cannot leave, as it is subjected to an infinite inwards acceleration, and the dynamical equations does not apply anymore: the field degrees of freedom are now doomed, translations in the direction perpendicular and tangent to the surface are not physical, the former because the system is trapped, and the latter because there are no dynamical laws at the surface.

One way to understand this fact is through inspection of the constraints
\begin{equation}
    G_1=p_\phi-3a^3\lambda_2\Delta^2\rho\approx0,\hspace{5mm}G_2=p_\rho\approx0.
\end{equation}
The Poisson brackets between them are $\{G_1,G_2\}=-a^3\varepsilon_{,X}=-3a^3\lambda_2\Delta(5X-1)$. Thence, when the surface is reached $\{G_1,G_2\}=0$, the constraints go from second to first class and the degrees of freedom of the system apparently become gauge. Notwithstanding, to be trapped in the surface yields a new constraint
\begin{equation}
    \varphi=\rho-\rho_{\pm}\approx0,
\end{equation}
whose Poisson brackets are $\{\varphi,G_1\}=0$ and $\{\varphi,G_2\}=1$. Therefore, only $G_1$ remains a first class constraint, a generator of gauge transformations in the direction tangent to the degeneracy surface, whereas $G_2$ is still second class. We say that the gauge in the perpendicular direction is fixed by the condition that the system remains in the surface \cite{STZ1,ANZ1}.

As a last remark, note that the Legendre transformation to write the first order Lagragian \eqref{L1} from \eqref{L2} is not globally invertible. In fact, invertibility fails at the degeneracy surface. This could sound as a problem, as the theories would not be equivalent. However, both Lagrangians describe identically the system in the non-degenerate regions, where the dynamical laws are equivalent. The conclusion drawn from both Lagrangians is the same: the system that reaches the degeneracy is trapped on the degenerate surface, where the equations of motion for the degrees of freedom involved in this evolution are no longer valid.

\section{Perturbations}  
\label{sec5}

Previous cosmological time crystals proposals (degenerate k-essence fields in our interpretation) necessarily violate the NEC, either becoming unstable or being salvaged by free time-dependent parameters from an effective field theory approach. Here we propose a model that reaches the degeneracy surface with $c^2_s=0$, satisfying the NEC and avoiding gradient instabilities. Nonetheless, the degeneracy is in some sense an extreme surface of phase space, so that naturally comes the question of whether other inconsistencies appear as it is reached. For that, we must investigate the behavior of perturbations in order to assess the validity of the model as it degenerates.

Consider the Mukhanov-Sasaki equation describing perturbations in the scalar field and in the scalar degrees of freedom of the metric \cite{GM1},
\begin{equation}
    v''_k+\left(\frac{c^2_sk^2}{\lambda_1^8}-\frac{z''}{z}\right)v_k=0,
\end{equation}
with $z=a^2\sqrt{2X\varepsilon_{,X}}/\lambda^4_1\mathcal{H}$, the slash denotes derivatives w.r.t. the conformal time $\eta=\int\rmd t/a$. When the degeneracy is reached, $z\propto\sqrt{\Delta}\rightarrow0$ and $z''\propto\Delta^{-7/2}\rightarrow\infty$, while $c^2_s\rightarrow0$, therefore, we are in the long wavelength limit, i.e.  $c^2_sk^2/\lambda^8_1\ll z''/z$, so
\begin{equation}
    v_k=C_1(k) z+C_2(k)z\int \frac{\textrm{d}\eta}{z^2}.
\end{equation}
To assert the health of the model we shall investigate the behaviour of the curvature perturbations and the Bardeen potential
\begin{align}
    \zeta_k&=\frac{v_k}{z}=C_1(k) +C_2(k)\int \frac{\textrm{d}\eta}{z^2},\\
    \Phi_k&=-\frac{3\mathcal{H}}{a\lambda^4_1 k^2}z^2\zeta'_k=-\frac{3\mathcal{H}}{a\lambda^2_1k^2}C_2(k).
\end{align}
The latter is well behaved as the system degenerates, yet, the integral on the second term in $\zeta_k$
\begin{equation} \label{d-eta/zeta2}
    \int_{\eta_i}^{\eta_d} \frac{\textrm{d}\eta}{z^2}=\int_{z_i}^{0} \frac{\textrm{d}z}{z'z^2},
\end{equation}
can be a problem, as, apparently, the end limit of the integral is divergent. Still,
\begin{multline}
    \frac{z'}{\lambda^4_1}=\sqrt{2X\varepsilon_{,X}}\left(\frac{2aa'}{\mathcal{H}}-\frac{\mathcal{H'}a^2}{\mathcal{H}^2}\right)+\\\frac{a^2X'}{\sqrt{2X\varepsilon_{,X}}\mathcal{H}}\left(\varepsilon_{,X}+\varepsilon_{,XX}X\right),
\end{multline}
and one can see from Eq. (\ref{eq10}) that $a$ and its derivatives are well behaved while, from Eq. (\ref{eq11}), 
\begin{equation}
   X'\approx-\frac{a\sqrt{2X}V_{,\phi}}{\varepsilon_{,X}},
\end{equation}
as $\Delta\rightarrow0$. Therefore, $z$ approaches 0 infinitely fast, as $z'\propto\Delta^{-3/2}\rightarrow\infty$, and the upper limit of the integral is well behaved: $(z'z^2)^{-1}\propto\Delta^{1/2}\rightarrow0$.

Thence, the gauge independent perturbations freezes into finite constants, just as the scalar field. In fact, in Ref. \cite{EV1}, it is argued that higher order quantum corrections should be taken into account, as the measure of the action $z$ would go to zero, yet, we have shown that this just means that the perturbations freeze into a finite non zero value (remaining in the linear regime), and have no dynamics anymore.

Accordingly, the overall behavior of the perturbations will be in general the same as any \kit-essence theory with oscillating modes that cross the sound horizon, become constant as $\zeta_k=C_1(k)$, and remaining approximately constant until the system freezes in the degeneracy (the decaying modes do not contribute, as the upper limit in the integral \eqref{d-eta/zeta2} tends to zero). Henceforth, the power spectrum for scalar perturbations will be
\begin{equation}
    P^{\zeta}_k=64\left.\frac{\mu^4_d\kappa^4\varepsilon}{c_s(1+p/\varepsilon)}\vline\right._{S},
\end{equation}
the subscript s denotes that the quantities are taken at the sound horizon crossing \cite{GM1}.

Thence, the phenomenology of inflation is the same. Notwithstanding, the parameters are now constrained not only by the perturbations, but also by the requirement that they source the present accelerated expansion.
%

\section{Final remarks}  
\label{sec6}

The degeneracy greatly enriches the dynamical features of non-linear systems, providing an interesting dynamical dimensional reduction process that divides the phase space into disjoint regions. In this proposal, it is throughout this mechanism that the \kit--essence field not only becomes a cosmological constant and drives the current accelerated phase of the Universe, but it conducts an earlier slow roll inflationary cosmological evolution as well. Also, the bounded dynamics eliminates the drawbacks previously found for such systems: regions which violate the NEC, perturbations that could violate causality, system instabilities, Hamiltonian unbounded from below, are all avoided.

The theory has two attractors, one originated from the potential term, the other coming from the non-canonical kinetic term. They are somewhat independent: the kinetic term is negligible in the slow rolling era, while being the responsible for the late time dynamics. The parameters have a disturbing scale difference: a very big effective primordial cosmological constant, and a much smaller one at late times. In this proposal, the potential is necessary and defines the value of the degenerate field, linking both phases of accelerated expansion, turning the theory more predictive. Along with that, the model allows an interesting singularity-free Universe, which starts in a de Sitter unstable equilibrium near the repulsive part of the degeneracy surface, for an unspecified large value of $|\phi|$, transits to a primordial slow roll inflationary phase with a possible subsequent standard FRWL evolution, and ends in a late time de Sitter evolution with a much smaller cosmological constant. 

The model can be viewed as a possible realization of  Penrose's Conformal Cyclic Cosmology (CCC) proposal \cite{P1}, in which a de Sitter final phase is connected through a conformal symmetry to a subsequent de Sitter beginning, and a new cycle of the universe emerges. Note that in the final phase of our model, the scalar field falls into the degeneracy surface where its amplitude is not a physical degree of freedom anymore, and it may possibly jump to the unstable part of the degeneracy surface, perhaps mediated by a conformal symmetry generator, or some other mechanism (quantum fluctuations?), initiating a new cycle. This is something to be explored in future investigations

Note that such cyclic Universes, as well as the model of Ref.~\cite{anna}, in which there is a net growth of the scale factor, have, however, been shown to be geodesic past--incomplete \cite{KS1,KMS1} and, therefore, not truly cyclical, as the geodesics meet somewhere in the past. The geodesic completeness of the present model will be addressed in future investigations.
This paper is a first try to show that \kit--essence fields can degenerate without instabilities, produce both phases of accelerated expansion in our Universe, and, surprisingly, a curvature singularity free model. For that, we started with a rather simple model in which space-time is filled solely with the \kit--essence field. Nevertheless, bringing more reality to the model may possibly ease some discomforts presented --the small value needed for the parameters in order to accomplish realistically both accelerated phases. The presence of other fluids dominating after inflation could make the Hubble drag decay slower, bringing the field to degenerate with smaller values; also, the $\phi$ dependence of $\lambda_2(\phi)$ can be used either to find a scaling solution in the presence of other fields, or to model a dynamical decay on the value of $\mu_f$ as the degeneracy is approached; finally, a reheating mechanism, in which the field thermalizes the Universe after primordial inflation ends, can be introduced through the interaction of the scalar with other fundamental fields, making it lose energy and fall into the degeneracy surface with much less energy than that obtained here, without the need for very small parameters.

Finally, it is important to stress that such modifications are not likely to change much the overall picture. During inflation, the energy density of the field $\phi$ is much larger than the contribution of the rest, and the process will be roughly the same as described here. After that, with the degeneracy at $X-1=0$, if in the beginning $X>1$, $X$ inevitably decays and $\phi$ becomes a cosmological constant. Also, $H\Delta^2$ and $\frac{d\lambda_2}{d\phi}\Delta^2$ in eq. (\ref{eq11}) do not contribute near the degeneracy, and will not modify its character. The only appreciable change will be in the relation between the parameters in the Lagrangian and the values the inflationary solutions approach the degeneracy.
Summing up, we have constructed a stable cosmic time crystal, and shown that interpreting it as degenerate systems can lead to improvements, alleviating the problems due to the highly non-linear dynamics, and allowing just one field to play the role of the inflaton and of the dark energy today, turning the theory more economical and predictive, while shedding some new light into this new dynamical dimensional reduction process: the degeneracy.

\begin{acknowledgments}
ALFJ was partially funded by the Research Support Foundation of Esp\'{\i}rito Santo - FAPES grant number 13/2019 and by CNPq of Brazil under grant number 200949/2022-5. NPN acknowledges the support of CNPq of Brazil under grant PQ-IB 310121/2021-3. JZ's research was partially supported by grants 1201208 and 1220862 from FONDECYT.
\end{acknowledgments}


\end{document}